\documentclass[aps,twocolumn,showpacs]{revtex4}
\usepackage{graphicx}

\begin{document}
\title{Comment on "Magnetoplasmons in a 2D electron fluid: Disk geometry".}
\author{M.V. Cheremisin}
\affiliation{A.F.Ioffe Physical-Technical Institute, St.Petersburg, Russia}
\date{\today}
\begin{abstract}
The main results of the commented article do not raise objections, but their final processing seems doubtful. Namely, the spectra of zero-field plasmon  with a fixed angular momentum $l = 0.1..4$ was scaled using a certain reference frequency and, then displayed as
a function of a specific screening parameter. Oddly enough, it turned out that the reference frequency includes the screening parameter explicitly. 
Obviously, it is a trivial error in scaling procedure. For this reason the physical meaning of the result becomes lost,
especially for zero value of screening parameter. The similar scaling method was performed for magnetoplasmon spectra at fixed values $0;0.5;\infty$ of screening parameter. Again, the case of zero screening parameter is doubtful. Both examples are exactly the source of misunderstandings. The original data for the axisymmetric plasmon $l=0$ were corrected and, then compared with the that followed from so-called telegraph line model, showing a difference of only 5 percent. We claim that the telegraph line model would be useful for describing disk-shaped plasmon devices with/without ring gratings.
 \end{abstract}
\maketitle

Fetter investigated\cite{Fetter86} the magnetoplasmon excitations for gated two-dimensional electron gas(2DEG) disk of a fixed radius $R$. Assuming that carriers behave as a fluid the original hydrodynamic approach was developed. The linearized Euler's and continuity equations were solved jointly with non-retarded Poisson equation. Finally, the spectrum of magnetoplasmon excitation was obtained. Although the approach looks solid, the presentation of the basic results seems  doubtful.

\begin{figure}[tbp]
\begin{center}\leavevmode
\includegraphics[width=1.0\linewidth]{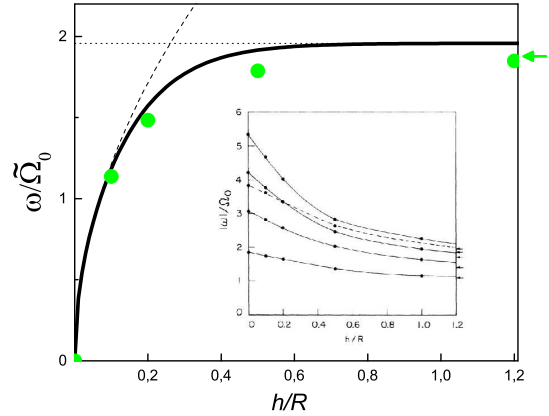} \caption[]{\label{Fig} Inset: the zero-field plasmon spectra at $l=1,2,3,4$ and $l=0$(dashed curve) reproduced from Ref.\cite{Fetter86}. Arrows at the right depict the frequencies for $l=1,2,3,0,4$ modes expected for unscreened 2D gas when $\Delta \rightarrow \infty$. Main panel: the dependence of the dimensionless frequency $\omega/\tilde{\Omega}_{0}$ vs screening parameter $\Delta$ specified by Eq.(\ref{Fetter}) for axisymmetric plasmon($l=0$). Thin dashed and dotted lines represent the asymptote $\sim \alpha_{02}\sqrt{\Delta}$ for highly screened and $\sqrt{\alpha_{02}}=1.957$ for unscreened 2DEG respectively. The data points and arrow mark corresponds to those depicted in inset for axisymmetric plasmon.}
\end{center}
\end{figure}

To confirm our conclusion, one may notice somewhat unusual configuration\cite{Fetter86} of 2DEG layer surrounded by pair of infinite grounded gates located parallel to disk at a distance $\pm h$. Then, two gaps between the gate and the layer were assumed to be filled by a material with a dielectric constant $\varepsilon$ and vacuum respectively. The author\cite{Fetter86} introduced the key parameter $\Delta=h/R$ responsible for 2D carrier screening. In Fig.\ref{Fig},inset we reproduce the result of the commented paper for zero-field dimensionless frequency $\omega/\Omega_{0}$ vs screening parameter $\Delta$ for $l=0,1..4$ angular momentum modes. One can pay attention to the author's choice of the reference frequency $\Omega_{0}=\sqrt \frac{4\pi e^{2}n \tanh{\Delta}}{m(1+\varepsilon)R}$, where $m$ and $n$ determine the carrier effective mass and density respectively. Emphasizing that $\Omega_{0}$ contains the screening parameter $\Delta$ explicitly, we come to a conclusion that the resulting plot in Fig.\ref{Fig},inset has no practical meaning. Indeed, the picture may mislead an inexperienced reader due to the apparent increase in frequency appeared for  enhanced screening at $\Delta\rightarrow 0$.

To resolve the problem, it is convenient to introduce modified reference frequency $\tilde{\Omega}_{0}=\Omega_{0}/\sqrt{\tanh{\Delta}}$ relevant for zero-field plasmon in ungated\cite{Stern67} 2D system. Consequently, we re-plot in Fig.\ref{Fig} the original data points attributed to axisymmetric mode($l=0$) in terms of renewed dimensionless frequency $\omega/\tilde{\Omega}_{0}$. As expected, the stronger the carrier screening the lower the plasmon frequency giving a clear physical sense.

Note that Fetter's result\cite{Fetter86} is accurate because it follows from the solution of linearized Euler, continuity and Poisson equations decomposed by cylindrical Hankel functions. Fortunately, the present case of zero angular momentum axisymmetric mode can be easily treated within so-called transmission line(TL) approach\cite{Aizin12}. Actually, TL approach represents a truncated Fetter's model adapted for 2DEG confined in a narrow stripe. It is noteworthy that the TL model implies a simplified solution of the Poisson equation, which gives a local dependence between the density of 2D carriers and in-plane potential. With this simplification, the TL approach yields equations similar to those known as telegraphist equations. The plasma wave frequency can be readily found in terms of inductance and capacitance per unit length of 2D stripe. Recently, the TL method was probed\cite{Cheremisin23} for axisymmetric plasmon description emphasizing that all variables depend only on the radius. Following the reasoning raised in Ref.\cite{Cheremisin23} and, moreover, accounting the actual sample geometry\cite{Fetter86} the zero-field axisymmetric mode frequency yields
\begin{equation}
\omega/\tilde{\Omega}_{0}=\sqrt{\alpha_{0m} \tanh(\alpha_{0m}\Delta)}.
\label{Fetter}\\
\end{equation}
Here, $\alpha_{0m}$ is the mth zero of the Bessel's function derivative $J'_{0}(x)$. For fundamental plasmon mode one put $\alpha_{02}=3.831$. The dependence specified by Eq.(\ref{Fetter}) is plotted in Fig.\ref{Fig}. Fetter's result coincides with that provided by TL model for highly screened disk $\Delta<1$ since Poisson equation solution remains local in both cases. In the opposite ungated disk case $\Delta>1$ TL approach predicts a higher value $\sqrt{\alpha_{02}}=1.957$ compared to exact one $1.861$ reported in Ref.\cite{Fetter86}. We refer this discrepancy to simplified solution of the Poisson equation used in TL approach. Noticing the difference between both results is as lower as 5 percent we conclude the TL model would be useful for spectral analysis\cite{Cheremisin23} of the disk shaped plasmonic devices with/without ring gratings.

To conclude, a data scaling error was revealed in the commented article. The correct data treatment for particular case of axisymmetric plasmon mode is provided. We justify the applicability of the transmission line model for disk-shaped plasmonic devices with/without ring gratings. Unfortunately, the comment was found not to meet the ultra-high criteria of the original article journal.

\bibliography{Comment}

\end{document}